\documentclass{appolb}

\usepackage{amssymb}
\usepackage{psfrag}

\usepackage{epsfig}
\usepackage{epsf}

\usepackage{amsmath}
\usepackage{amsfonts}
\usepackage{psfrag,epsfig,graphicx}

\usepackage{graphicx}

\newcommand{\pom}{\mathbb{P}}

\def\runhead{\textsc{
Photon dissociation into two and three jets}}

\begin{document}
\title{
Photon dissociation into two and three jets:
\\
initial and final state corrections 
\thanks{Presented at the 16th conference on Elastic and Diffractive Scattering (EDS Blois 2015).}
}
\author{R. Boussarie$^{1}$, A. Grabovsky$^{2,3}$, L. Szymanowski$^{4}$, S. Wallon$^{1,5}$
\address{$^{1}$
Laboratoire de Physique Th\'{e}orique, UMR 8627, CNRS, Univ. Paris Sud, Universit\'{e} Paris Saclay, 91405 Orsay, France \\
\vspace{.2cm}
$^{2}$
Novosibirsk State University,
2 Pirogova street, Novosibirsk, Russia\\
\vspace{.2cm}
$^{3}$
Theory division, Budker Institute of Nuclear Physics,
11 Lavrenteva avenue, Novosibirsk, Russia
\\
\vspace{.2cm}
$^{4}$National Centre for Nuclear Research (NCBJ), Warsaw, Poland \\
\vspace{.2cm}
$^{5}$UPMC Universit\'{e} Paris 6, Facult\'{e} de physique, 4 place Jussieu, 75252 Paris Cedex 05, France
}
}
\maketitle

\begin{abstract}	
We study the impact factor for the photon to quark, antiquark and gluon transition within
Balitsky's shock-wave formalism. 
Our aim is to extend existing results
beyond approximations discussed in the literature.
We present our results of the real contribution, and present some intermediate results on virtual contributions for the photon to quark, antiquark transition. 

\end{abstract}

\section{Introduction}

Among the achievements of HERA, one of the major results was the experimental evidence \cite{Derrick:1993xh,Ahmed:1994nw}
that 
among the whole set of $\gamma^* p \to X$ deep inelastic scattering events, almost 10\%  are diffractive (DDIS), of the form $\gamma^* p \to X Y$ with a rapidity gap between the proton remnants $Y$
and the hadrons $X$
coming from the fragmentation region of the initial virtual photon.

\begin{figure}[h]
\center
\psfrag{q}{\raisebox{-.2cm}{$\gamma^*$}}
\psfrag{l1}{$e^\pm$}
\psfrag{l2}{$e^\pm$}
\psfrag{P}{$\pom$}
\psfrag{ld}{}
\psfrag{lu}{}
\psfrag{R}{}
\psfrag{q1}{\raisebox{.2cm}{\ \ jet}}
\psfrag{q2}{\raisebox{-.3cm}{\ \ jet}}
\psfrag{p1}{$p$}
\psfrag{p2}{$Y$}
\includegraphics[scale=.90]{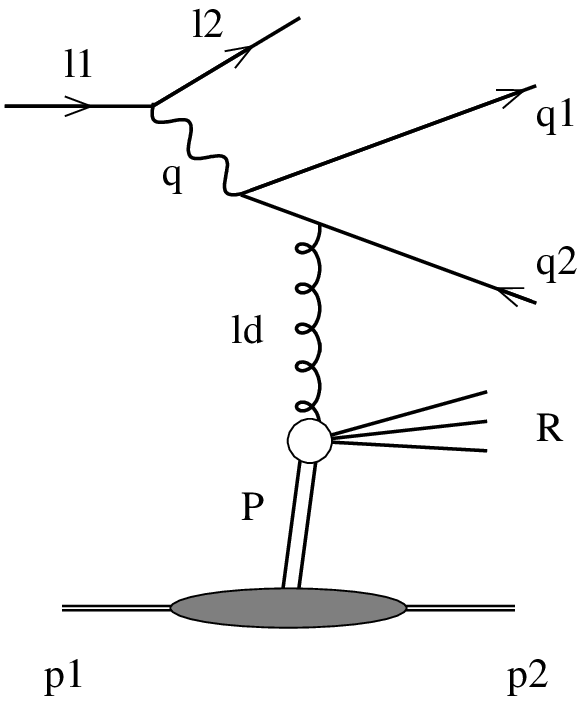}
\qquad 
\psfrag{q2}{\raisebox{-.4cm}{\ \ jet}}
\raisebox{.5cm}{\includegraphics[scale=.90]{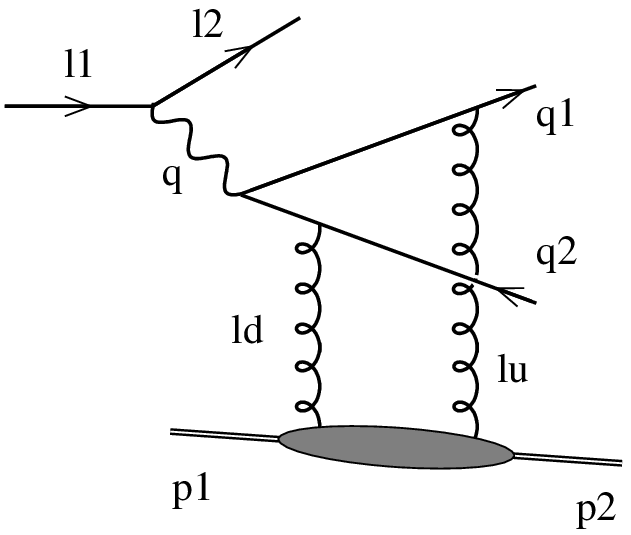}}
\caption{Resolved (left panel) and direct Pomeron (right panel) contributions to two jets production.}
\label{ResDirect}
\end{figure}

There are two main approaches to theoretically describe 
diffraction. The first one involves
a {\em resolved} Pomeron contribution, see Fig.~\ref{ResDirect} (left),  while the second one
relies on a  {\em direct} Pomeron contribution involving the coupling of a Pomeron with the diffractive state, see Fig.~\ref{ResDirect} (right). 

For moderate invariant mass $M^2$ of the  diffractively produced state $X$, such a state
 can be modeled in perturbation theory by a  $q \bar{q}$ pair,  or by higher Fock states as a $q \bar{q} g$ state for larger values of $M^2$. Based on such a model, with
 a two-gluon exchange picture for the Pomeron,  a good description of HERA data for diffraction~\cite{Chekanov:2004hy-Chekanov:2005vv-Chekanov:2008fh,
Aktas:2006hx-Aktas:2006hy-Aaron:2010aa-Aaron:2012ad-Aaron:2012hua} could be achieved~\cite{Bartels:1998ea}. 
In the direct components considered there, the $q \bar{q} g$ diffractive state has been studied in two particular limits. The first one, valid for very large $Q^2$, corresponds to a collinear approximation in which the transverse momentum of the gluon is assumed to be much smaller than the transverse momentum of the emitter~\cite{Wusthoff:1995hd-Wusthoff:1997fz}. 
The second one~\cite{Bartels:1999tn,Bartels:2002ri}, valid for very large $M^2$, is based on the assumption of a strong ordering of longitudinal momenta, encountered in BFKL equation~\cite{Fadin:1975cb-Kuraev:1976ge-Kuraev:1977fs-Balitsky:1978ic}. Both these approaches were combined in order to describe HERA data for DDIS~\cite{Marquet:2007nf}.

It would be natural to extend the HERA studies
to similar hard diffractive events at LHC. 
The idea here is to adapt the concept of  photoproduction of diffractive jets, which  was performed at HERA~\cite{Chekanov:2007rh,Aaron:2010su}, now with a flux of
quasi-real photons in ultraperipheral collisions (UPC)~\cite{Baltz:2007kq-Baur:2001jj}, relying on the notion of equivalent photon approximation. In both cases, 
 the hard scale is provided by the invariant mass of the tagged jets.

We here report on our computation~\cite{Boussarie:2014lxa} of the $\gamma^* \to q \bar{q} g$ impact factor at tree level with an arbitrary number of $t$-channel gluons described within the Wilson line formalism, also called QCD shockwave approach~\cite{Balitsky:1995ub-Balitsky:1998kc-Balitsky:1998ya-Balitsky:2001re}. As an aside, we rederive the $\gamma^* \to q \bar{q}$ impact factor. In particular, the 
$\gamma^* \to q \bar{q} g$ transition is computed without any soft or collinear approximation for the emitted gluon, in contrast with the above mentioned calculations. These results provide a necessary generalization of building blocks for inclusive DDIS (of potential significant phenomenological importance~\cite{Motyka:2012ty}) as well as for two- and three-jet diffractive production.

\section{The shockwave formalism in a nutshell}

Balitsky's shockwave formalism
is very powerful in determining evolution equations and impact factors at next-to-leading order for inclusive processes~\cite{Balitsky:2010ze-Balitsky:2012bs}, at semi-inclusive level for $p_t$-broadening in $pA$ collisions~\cite{Chirilli:2011km-Chirilli:2012jd} or in the evaluation of the triple Pomeron vertex beyond the planar limit~\cite{Chirilli:2010mw}, when compared with usual methods based on summation of contributions of individual Feynman diagrams computed in momentum space. It is an effective way of estimating the effect of multigluon exchange, formulated in coordinate space and thus natural in view of describing saturation~\cite{GolecBiernat:1998js-GolecBiernat:1999qd}. 

We introduce the light cone vectors
$n_{1}$ and $n_{2}$%
\begin{equation}
\label{Sudakov-basis}
n_{1}=\left(  1,0,0,1\right)  ,\quad n_{2}=\frac{1}{2}\left(  1,0,0,-1\right)
,\quad n_{1}^{+}=n_{2}^{-}=n_{1} \cdot n_{2}=1 \,,
\end{equation}
and the Wilson lines as 
\begin{equation}
U_{i}=U_{\vec{z}_{i}}=U\left(  \vec{z}_{i},\eta\right)  =P \exp\left[{ig\int_{-\infty
}^{+\infty}b_{\eta}^{-}(z_{i}^{+},\vec{z}_{i}) \, dz_{i}^{+}}\right]\,.
\label{WL}%
\end{equation}
The operator $b_{\eta}^{-}$ is the external shock-wave field built from slow gluons 
whose momenta are limited by the longitudinal cut-off defined by the rapidity $\eta$
\begin{equation}
b_{\eta}^{-}=\int\frac{d^{4}p}{\left(  2\pi\right)  ^{4}}e^{-ip \cdot z}b^{-}\left(
p\right)  \theta\left(e^{\eta}-\frac{|p^{+}|}{P^+}\right)\,,\label{cutoff}%
\end{equation}
where $P^+$ is the typical large $+$ momentum of the problem, to be identified with $p_\gamma^+$ later on. We will denote the longitudinal cut-off $\sigma = e^\eta \, P^+ = \alpha \, P^+.$

We use the light cone gauge
$\mathcal{A}\cdot n_{2}=0,$
with $\mathcal{A}$ being the sum of the external field $b$ and the quantum field
$A$%
\begin{equation}
\mathcal{A}^{\mu} = A^{\mu}+b^{\mu},\;\;\;\;\;\;\;\;\;\quad b^{\mu}\left(  z\right)  =b^{-}(z^{+},\vec{z}\,) \,n_{2}%
^{\mu}=\delta(z^{+})B\left(  \vec{z}\,\right)  n_{2}^{\mu}\,,\label{b}%
\end{equation}
where
$B(\vec{z})$ is a profile function. 
Indeed, let us consider an external gluon field $b^{\mu}$ in its rest frame and boost it along the $+$ direction. One obtains :
\begin{eqnarray}\nonumber
&&b^+ \! \left( x^+,\, x^-, \, \vec{x} \right) \rightarrow \frac{1}{\lambda}b^+ \left( \lambda x^+,\, \frac{1}{\lambda} x^- ,\, \vec{x} \right)\,, \\ \nonumber
&&b^- \! \left( x^+,\, x^-, \, \vec{x} \right) \rightarrow {\lambda} b^- \left( {\lambda x^+},\, {\frac{1}{\lambda} x^-} ,\, \vec{x} \right)\,, \\ \nonumber
&&b^i \, \left( x^+,\, x^-, \, \vec{x} \right) \rightarrow \, \, \, b^i \, \left( \lambda x^+,\, \frac{1}{\lambda} x^- ,\, \vec{x} \right)\,. \\ \nonumber
\end{eqnarray}
Assuming that the field vanishes at infinity, one immediately gets that only its minus component survives the boost in the limit $\lambda \to \infty\,,$ and that it does not depend on $x^-$ and contains $\delta \left( x^+ \right)\,,$ thus justifying the form of $b^\mu$ in Eq.~(\ref{b}). 

We use intensively in the following the dipole operator 
 constructed from the Wilson line (\ref{WL}), namely
 $\mathbf{U}_{12}=\frac{1}{N_{c}}\rm{tr}\left(  U_{1}U_{2}^{\dagger}\right)  -1\,.$

\section{Impact factor for $\gamma\rightarrow q\bar{q}$ transition}

\begin{figure}
\center
\includegraphics[scale=0.65]{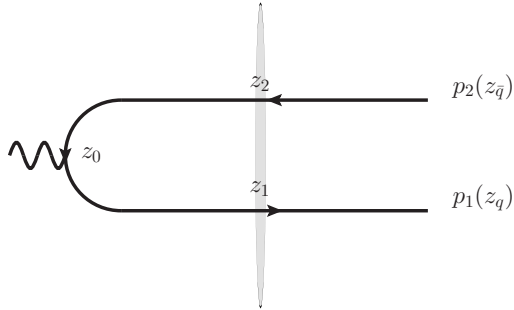}
\caption{Diagram contributing to the impact factor for two jet production }
\label{leading}
\end{figure}

At leading order, the diagram contributing to the impact factor  for $\gamma\rightarrow q\bar{q}$ transition is shown in Fig. \ref{leading}, in which $z's$ denote the 
coordinates of interaction points with the photon and the shock wave.
After projection on the color singlet state and subtraction of the contribution without interaction with the shock wave, the contribution of this diagram can be written in the momentum space as (factorizing out a global QED factor $-i e_q$)
\begin{equation}
M_{0}^{\alpha}=N_c \int d\vec{z}_{1}d\vec{z}_{2}F\left(  p_{q},p_{\bar{q}}%
,z_{0},\vec{z}_{1},\vec{z}_{2}\right)  ^{\alpha} \mathbf{U}_{12}\,.
\label{M0int}%
\end{equation}
Denoting $Z_{12} = \sqrt{x_{q}x_{\bar{q}}\vec{z}_{12}^{\,\,2}}$, we get for a longitudinally polarized photon
\begin{eqnarray}
\label{FL}
F\left(  p_{q},p_{\bar{q}},k,\vec{z}_{1},\vec{z}_{2}\right)  ^{\alpha
}\varepsilon_{L\alpha}&=&\theta(p_{q}^{+})\,\theta(p_{\bar{q}}^{+})\frac
{\delta\left(  k^{+}-p_{q}^{+}-p_{\bar{q}}^{+}\right)  }{(2\pi)^{2}}%
e^{-i\vec{p}_{q}\cdot \vec{z}_{1}-i\vec{p}_{_{\bar{q}}}\cdot\vec{z}_{2}}
\nonumber \\
&\times&
(-2i)\delta_{\lambda_{q},-\lambda_{\bar{q}}}\,x_{q}x_{\bar{q}}%
\,Q\,K_{0}\left(Q \, Z_{12}\right)\,,
\end{eqnarray}
whereas for a transversally polarized photon
\begin{eqnarray}
\label{FT}
F(  p_{q},p_{\bar{q}},k,\vec{z}_{1},\vec{z}_{2})  ^{j}%
\varepsilon_{Tj}\!
&=&\theta(p_{q}^{+})\,\theta(p_{\bar{q}}^{+})\frac{\delta(
k^{+}\!\!-\!p_{q}^{+}\!-p_{\bar{q}}^{+}\!)  }{(2\pi)^{2}}e^{-i\vec{p}_{q}\cdot\vec
{z}_{1}-i\vec{p}_{_{\bar{q}}}\cdot\vec{z}_{2}}
\nonumber \\
&&\hspace{-2cm}\times
\delta_{\lambda_{q},-\lambda_{\bar{q}}}( x_{q}-x_{\bar{q}%
}+s\lambda_{q})  \frac{\vec{z}_{12} \cdot \vec{\varepsilon}_{T}}{\vec{z}_{12}^{\,\,2}}
Q \,Z_{12} K_{1}(Q\, Z_{12})\,.\!\!\!\!\!
\end{eqnarray}

\section{Impact factor for $\gamma\rightarrow q\bar{q}g$ transition}

\begin{figure}
\center
\includegraphics[scale=0.65]{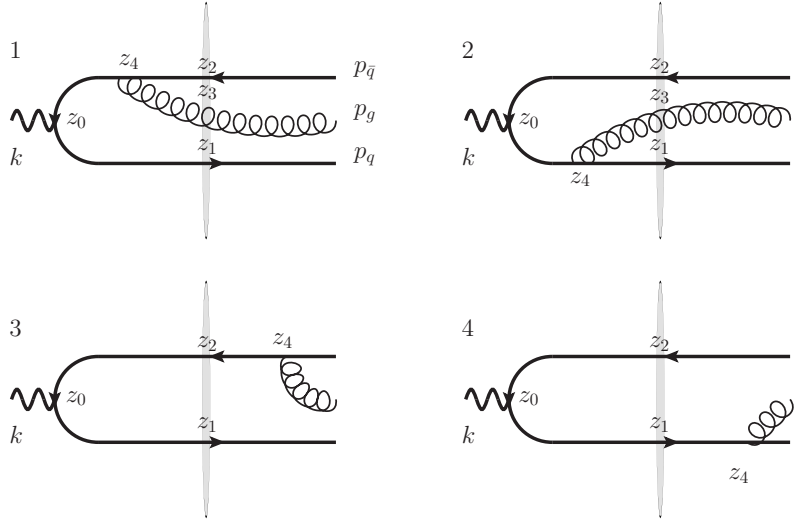}
\caption{Diagrams contributing to the impact factor for three jet production}
\label{3body}
\end{figure}

In the case of the $q\,\bar q\,g$ Fock final state the contributiong diagrams are shown in Fig.~\ref{3body}.
After projection on the color singlet state and subtraction of the contribution without interaction with the shock wave,  the result can be put in the form 
\begin{eqnarray}
\nonumber 
M^{\alpha} &=& N_c^2 \int d\vec{z}_{1}d\vec{z}_{2}d\vec{z}_{3} \, F_{1}\left(  p_{q},p_{\bar{q}}%
,p_{g},z_{0},\vec{z}_{1},\vec{z}_{2},\vec{z}_{3}\right)  ^{\alpha} \nonumber \\
&\times& \frac{1}{2}
\left( \mathbf{U}_{32} + \mathbf{U}_{13} - \mathbf{U}_{12} + \mathbf{U}_{32}\,\mathbf{U}_{13} \right)\nonumber
\\
&+& N_c \int d\vec{z}_{1}d\vec{z}_{2} \, F_{2}\left(  p_{q},p_{\bar{q}},p_{g},z_{0}%
,\vec{z}_{1},\vec{z}_{2}\right)  ^{\alpha}\frac{N_{c}^{2}-1}{2N_{c}} \mathbf{U}_{12}\,.
\label{F2tilde}%
\end{eqnarray}
In this equation, the first two lines and the third one correspond to contributions to the impact factor, respectively, of the diagrams 1 and 2  of      Fig.~\ref{3body} and of 
the diagrams 3 and 4 of it. The explicit expressions
for the functions $F_i,$
for both longitudinally and transversally polarized photon can be found in 
 ref.~\cite{Boussarie:2014lxa}.

\section{The 2- and 3-gluon approximation}

We first notice that the dipole operator $\mathbf{U}_{ij}$ involves terms at least of order $g^2$. Hence for only two or three exchanged gluons one can neglect the quadrupole term in the amplitude $M^{\alpha}$ which results in the simpler expression
\begin{eqnarray}
\label{M3gBis}
&& \hspace{-.3cm}M^{\alpha} \overset{\mathrm{g^3}}{=}   \frac{1}{2}\int d\vec{z}_{1}d\vec{z}%
_{2} \mathbf{U}_{12}  \left[ \left(  N_{c}^{2}-1\right)
\tilde{F}_{2}\left(  \vec{z}_{1},\vec{z}%
_{2}\right) ^{\alpha} \right.
\nonumber \\
&& \hspace{-.3cm}\left.
+ \int d\vec{z}_{3} \left\{  N_{c}^{2}F_{1}\left(
\vec{z}_{1},\vec{z}_{3},\vec{z}_{2}\right)^{\alpha} 
+N_{c}^{2}F_{1}\left(  \vec{z}_{3},\vec{z}%
_{2},\vec{z}_{1}\right)  ^{\alpha} -  F_{1}\left(  \vec{z}_{1},\vec{z}_{2},\vec{z}_{3}\right)  ^{\alpha} \right\} \right]\,.
\end{eqnarray}
Those integrals can be performed analytically
when $\vec{p}_q=\vec{p}_g=\vec{p}_{\bar{q}}=\vec{0}$. They are
otherwise expressible as a simple convergent integral over the interval $[0,1]$. 

\section{Towards the next-to-leading-order corrections}

\begin{figure}[h]
\centerline{\includegraphics[scale=0.75]{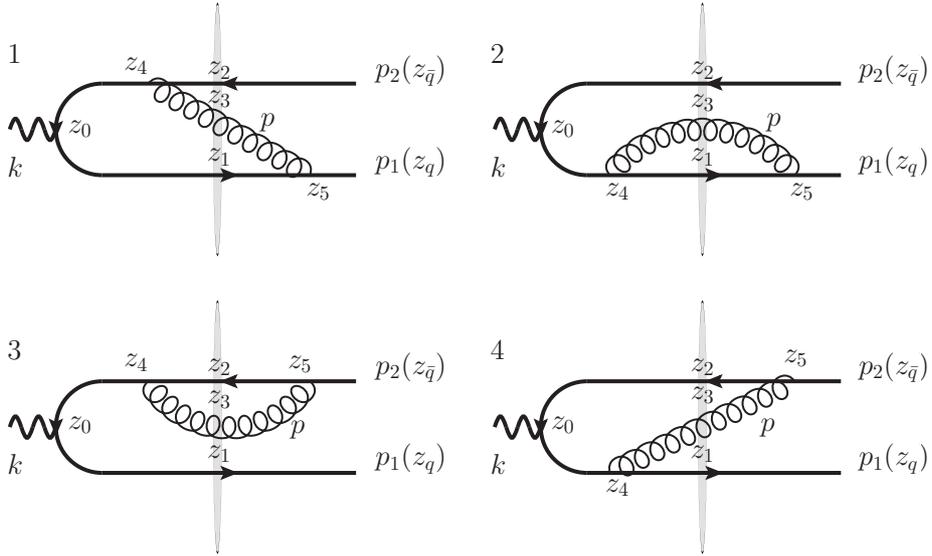}}
\caption{Diagrams contributing to virtual corrections in which the radiated gluon interacts with the shock wave.}
\label{nloSW}
\end{figure}

The virtual corrections to the $\gamma^* \to q \bar{q}$ involve two kinds of contributions.
\begin{figure}[h]
\centerline{\includegraphics[scale=0.65]{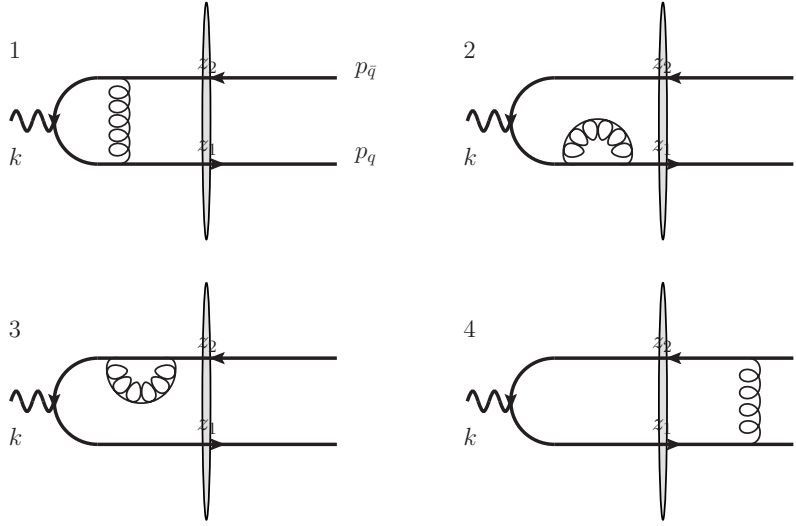}}
\caption{Diagrams contributing to virtual corrections in which the radiated gluon doesn't cross the shock wave.}
\label{nlo}
\end{figure}
\noindent
The diagrams contributing to virtual corrections in which the radiated gluon does not cross the shock wave are shown in Fig.~\ref{nlo},
and the diagrams in which the radiated gluon interacts with  the shock wave are illustrated in the Fig.~\ref{nloSW}.
One should note that although these virtual corrections only involve one-loop diagrams, the complications arise due to the presence of many different scales. Indeed, our aim is to obtain results in the general kinematics where the virtuality of incoming photon, the $t-$channel momentum transfer and the invariant mass $M^2$ of the diffractive two-jet state are arbitrary. Additionally, this impact factor  is a function of the virtuality of $t-$channel exchanged gluons. 

We now provide some intermediate results of our computation.
First, we present the matrix element corresponding to diagrams 1, 2, 3 of Fig.~\ref{nlo}~\cite{Boussarie:2015qet}.
We work in 
dimensional regularization for the transverse momentum space, i.e. $d=D-2 =2+ 2\,\epsilon\,,$ and introduce the regularization scale $\mu$, and the related dimensionless scale  $\tilde{\mu}^2=\mu^2/Q^2\,.$
Denoting $p_{ij} \equiv p_i-p_j\,,$ we introduce $p_\perp=p_{q1\perp}\,,$  $\vec{p}^{\,2}=-p_\perp^2$ and
$w=\vec{p}^{\,2}/Q^2\,.$ For simplicity, we write $x=x_q.$
We get for the case of a longitudinally polarized photon 
\begin{eqnarray}
&&\hspace{-1cm}T_{fi}|_{\epsilon_{\alpha}=n_{2\alpha}}=-i g^2\frac{N_{c}^{2}-1}{2N_{c}%
}tr(U(p_{1\bot})U^{\dag}(-p_{2\bot}))\delta(p_{q1\bot}-p_{\gamma\bot}%
+p_{\bar{q}\,2\bot})
\nonumber \\
&\times&\delta(p_{q}^{+}-p_{\gamma}^{+}+p_{\bar{q}}^{+})\theta
(p_{q}^{+})\theta(p_{\bar{q}}^{+})
\nonumber \\
&\times&\frac{\Gamma(1-\epsilon)}{\left(
16\pi^{3}\right)  ^{1+\epsilon}}\frac{1}{\sqrt{2p_{\gamma}^{+}}\sqrt
{2p_{q}^{+}}\sqrt{2p_{\bar{q}}^{+}}}
\frac{x(1-x)p_{\gamma}^{+}{}\overline{u}_{p_{q}}\gamma^{+}v_{p_{\bar{q}%
}}}{x(1-x)Q^{2}+\vec{p}{}^{\,\,2}} 
\nonumber \\
&\times& \left(  \left(  2\ln\left(  \frac
{(1-x)x}{\alpha^{2}}\right)  -3\right)  \left(  \ln\left(  \frac{\left(
w-x^{2}+x\right)  ^{2}}{(1-x)x\tilde{\mu}^{2}}\right)  +\frac{1}{\epsilon
}\right)  \right. \nonumber \\
&&\left.+\ln^{2}\left(  \frac{x}{1-x}\right)  -\frac{\pi^{2}}{3}+6\right)  \,. 
\end{eqnarray}
Expanding the photon momentum in the Sudakov basis (\ref{Sudakov-basis}) as
\begin{equation}
p_\gamma = p_\gamma^+ \, n_1 - \frac{Q^2}{2 p_\gamma^+} \, n_2
\end{equation}
one can explicitly check the electromagnetic gauge invariance for this group of diagrams
since
\begin{equation}
T_{fi}|_{\epsilon_{\alpha}=n_{1\alpha}}=\frac{Q^{2}}{2p_{\gamma}^{+2}}%
T_{fi}|_{\epsilon_{\alpha}=n_{2\alpha}}\,.
\end{equation}
Similarly, for the case of a transversally polarized photon, one gets
\begin{eqnarray}
&& \hspace{-.3cm}T_{fi}|_{transverse}=-i g^2 \frac{N_{c}^{2}-1}{2N_{c}}tr(U(p_{1\bot})U^{\dag
}(-p_{2\bot}))\delta(p_{q1\bot}-p_{\gamma\bot}+p_{\bar{q}2\bot})
\nonumber \\
&&\hspace{-.4cm}\times \,\delta
(p_{q}^{+}-p_{\gamma}^{+}+p_{\bar{q}}^{+})\theta(p_{q}^{+})\theta(p_{\bar{q}%
}^{+})
\nonumber \\
&&\hspace{-.4cm}\times 
\frac{\Gamma(1-\epsilon)}{\left(  16\pi^{3}\right)  ^{1+\epsilon}%
}\frac{\epsilon_{i}}{\sqrt{2p_{\gamma}^{+}}\sqrt{2p_{q}^{+}}\sqrt{2p_{\bar{q}%
}^{+}}}  \frac{-\left(  \frac{1}{2}\overline{u}_{p_{q}}[\gamma^{i}\hat{p}_{\bot
}]\gamma^{+}v_{p_{\bar{q}}}+(2x-1)p^{i}\overline{u}_{p_{q}}\gamma
^{+}v_{p_{\bar{q}}}\right)  }{2(x(1-x)Q^{2}+\vec{p}{}^{\,\,2})}%
\nonumber \\
&&\hspace{-.4cm} \times\!\!\left[ \! \left(  2\ln\left(  \frac{(1-x)x}{\alpha^{2}}\right)  -3\right)\!\!
\left( \! \ln\left(  \frac{w-x^{2}+x}{\tilde{\mu}^{2}}\right)  +\frac
{(1-x)x\ln\left(  \frac{(1-x)x}{w-x^{2}+x}\right)  }{w}+\frac{1}{\epsilon
}\!\right)  
\right. \nonumber \\
&& \hspace{-.4cm}\left.
+\,\ln^{2}\left(  \frac{x}{1-x}\right)  -\frac{\pi^{2}}{3}+6\right]  .
\end{eqnarray}
Second, we present the singular part of  diagram 4 of fig.~\ref{nlo} involving final state interaction. 
The result for a longitudinally polarized photon reads
\begin{eqnarray}
\label{final-state-long}
&& T_{fi}|_{\epsilon_{\alpha}=n_{2\alpha}}=i\frac{N_{c}^{2}-1}{2N_{c}%
}tr(U(p_{1\bot})U^{\dag}(-p_{2\bot}))
\delta(p_{\gamma\bot}-p_{1q\bot
}-p_{2\bar{q}\bot})
\nonumber\\
&& \times
\delta(p_{\gamma}^{+}-p_{q}^{+}-p_{\bar{q}}^{+})
\frac{\Gamma(1-\varepsilon)}{\left(
16\pi^{3}\right) ^{1+\varepsilon}}
\frac
{1}{\sqrt{2p_{\gamma}^{+}}\sqrt{2p_{q}^{+}}\sqrt{2p_{\bar{q}}^{+}}}\nonumber\\
&& \times\left\{ \frac{(1-x)x\bar{u}_{p_{q}}\gamma^{+}v_{p_{\bar{q}}%
}p_{\gamma}^{+}}{\vec{p}^{\,\,2}+Q^{2}(1-x)x}\left[ \ln^{2}\left(
\frac{(1-x)x}{\alpha^{2}}\right) -\ln^{2}\left( \frac{1-x}{x}\right) \right. \right.\nonumber \\
&&\left. \left. + \, 2\ln\left( \frac{(1-x)x}{\alpha^{2}}\right) \left( \ln\left(
\frac{\left( \vec{p}^{\,\,2}+Q^{2}(1-x)x\right) ^{2}}{Q^{2}(x\vec{p}%
_{\bar{q}}-(1-x)\vec{p}_{q})^{2}}\right) +i\pi\right) \right] +C_{\Vert
}^{fs}\right\} ,\,\,\,\,\,
\end{eqnarray}
while for a transversally polarized photon we obtain
\begin{eqnarray}
\label{final-state-transverse}
&& \hspace{-.2cm} T_{fi}|_{transverse}=i\frac{N_{c}^{2}-1}{2N_{c}}tr(U(p_{1\bot})U^{\dag
}(-p_{2\bot}))
\delta(p_{\gamma\bot}-p_{1q\bot}-p_{2\bar{q}\bot
})\nonumber\\
&&\hspace{-.2cm}\times \, \delta(p_{\gamma}^{+}-p_{q}^{+}-p_{\bar{q}}^{+})
\frac{\Gamma(1-\varepsilon)}{\left( 16\pi^{3}\right)
^{1+\varepsilon}}
\frac{\epsilon_{i}}%
{\sqrt{2p_{\gamma}^{+}}\sqrt{2p_{q}^{+}}\sqrt{2p_{\bar{q}}^{+}}}\nonumber\\
&&\hspace{-.2cm} \times\left\{ -\frac{(2x-1)p_{\bot}^{i}\bar{u}_{p_{q}}\gamma^{+}%
v_{p_{\bar{q}}}+\frac{1}{2}\bar{u}_{p_{q}}\gamma^{+}[\gamma_{\bot}^{i}\hat
{p}_{\bot}]v_{p_{\bar{q}}}}{\left( Q^{2}(1-x)x+\vec{p}^{\,\,2}\right)
}\left[ \frac{1}{2}\ln^{2}\left( \frac{(1-x)x}{\alpha^{2}}\right)
\right. \right. \nonumber \\
&&\hspace{-.2cm} \left. \left. -\frac
{1}{2}\ln^{2}\left( \frac{x}{1-x}\right) +\ln\left( \frac{(1-x)x}%
{\alpha^{2}}\right) 
\left( \frac{Q^{2}(1-x)x}{\vec{p}^{\,\,2}}\ln\left(
\frac{Q^{2}(1-x)x}{Q^{2}(1-x)x+\vec{p}^{\,\,2}}\right)
\right.\right.\right.
\nonumber \\
&&\hspace{-.2cm} \left.\left.\left.
+\ln\left(
\frac{(1-x)x\left( Q^{2}(1-x)x+\vec{p}^{\,\,2}\right) }{(x\vec{p}_{\bar{q}%
}-(1-x)\vec{p}_{q})^{2}}\right) +i\pi\right) \right] +C_{\bot}%
^{fs}\right\} \,.
\end{eqnarray}
In eqs.~(\ref{final-state-long},\ref{final-state-transverse}), $C_{\Vert
}^{fs}$ and $C_{\bot}^{fs}$ are finite terms which are 
too lengthy to be written here.

\section{Conclusion}

Dijet production in DDIS at HERA was recently analyzed~\cite{Aaron:2011mp}. A precise comparison of 
dijet versus triple-jet production, which has not been  performed yet at HERA~\cite{Adloff:2000qi}, would be of much interest. Investigations of the azimuthal distribution of dijets in diffractive photoproduction performed by ZEUS~\cite{Guzik:2014iba} show sign of a possible need for a 2-gluon exchange model, which is part of the shock-wave mechanism. Our calculation could be used for phenomenological studies of those experimental results.
Complementary studies could be performed at LHC with UPC events. A full quantitative first principle analysis of this will be possible after completing our program of  computing virtual corrections to the $\gamma^* \rightarrow q\bar{q}$ impact factor~\cite{Boussarie:prep}, for which we have provided here intermediate results.

\section*{Acknowledgments}

We thank Ian Balitsky, Cyrille Marquet and St\'ephane Munier for discussions. Andrey~V.~Grabovsky acknowledges support of president scholarship 171.2015.2,
RFBR grant 13-02-01023, Dynasty foundation, Metchnikov grant and University Paris Sud. He 
is also
grateful to LPT Orsay for hospitality  while part of the
presented work was being done. Renaud Boussarie thanks RFBR for financial support
via grant 15-32-50219.
This work was partially supported by the ANR PARTONS (ANR-12-MONU-0008-01), the COPIN-IN2P3 Agreement and the Th\'eorie-LHC France Initiative. Lech Szymanowski was supported by a grant from the French Ambassy in Poland.

\end{document}